# High Throughput Neural Network based Embedded Streaming Multicore Processors

[1]Raqibul Hasan, [1]Tarek M. Taha, [1]Chris Yakopcic, and [2]David J. Mountain
[1]University of Dayton, Ohio, USA
[2]Laboratory for Physical Sciences (a Department of Defense research lab) MD, USA
Email: [1]{hasanm1, tarek.taha, cyakopcic1}@udayton.edu, [2]djmount@lps.umd.edu

*Abstract*— With power consumption becoming a critical processor design issue, specialized architectures for low power processing are becoming popular. Several studies have shown that neural networks can be used for signal processing and pattern recognition applications. This study examines the design of memristor based multicore neural processors that would be used primarily to process data directly from sensors. Additionally, we have examined the design of SRAM based neural processors for the same task. Full system evaluation of the multicore processors based on these specialized cores were performed taking I/O and routing circuits into consideration. The area and power benefits were compared with traditional multicore RISC processors. Our results show that the memristor based architectures can provide an energy efficiency between three and five orders of magnitude greater than that of RISC processors for the benchmarks examined.

## I. INTRODUCTION

GENERAL purpose computing systems are used for a large variety of applications. Extensive supports for flexibility in these systems limit their energy efficiencies. Reliability and power consumption are among the main obstacles to continued performance improvement of future multicore computing systems [1]. As a result, several research groups are investigating the design of energy efficient processors from different aspects. These include architectures for approximate computation utilizing dynamic voltage scaling technique, dynamic precision control, and inexact hardware [2,3]. Application specific architectures are also proposed for several application domains such as signal processing and video processing [4].

Interest in specialized architectures for accelerating neural networks has increased significantly because of their ability to reduce power, increase performance, and allow fault tolerant computing. Recently IBM has developed the TrueNorth chip [5] consisting of 4,096 neurosynaptic cores interconnected via an intra-chip network. The basic building block is a core, implementing 256 spiking neurons each having 256 inputs. Their synapse element is SRAM based and off-chip training is utilized. DaDianNao [6] is an accelerator for deep neural network (DNN) and convolutional neural network (CNN). In this system neuron synaptic weights are stored in eDRAM and later brought into the Neural Functional Unit for execution.

Several systems, including surveillance, self-driving cars, pattern recognition in cameras are based on image processing tasks. Neural networks are widely used for pattern recognition, signal processing, and image processing applications [7-9]. This paper examines neural network based processing systems where input data are coming from an imaging sensor chip. The sensor chip is stacked on top of the processing chip using 3D integration technology. This enables data transfer at low power, and high bandwidth.

Memristors [10,11] have received significant attention as a potential building block for neuromorphic systems [12,13]. In these systems memristors are used in a crossbar structure. Memristor devices in a crossbar structure can evaluate many multiply-add operations in parallel in the analog domain very efficiently (these are the dominant operations in neural networks). In this paper we examine the impact of using memristor crossbars for the synaptic array within the neural cores.

The memristor crossbar based multicore embedded neural architectures are designed for sensor data processing. We have compared the system level area and power benefits of the proposed systems over a SRAM based digital system and a traditional multicore RISC system. Our results indicate that the memristor based architectures can provide between three to five orders energy efficiency over RISC processors for the selected benchmarks. Additionally, they can be up to 400 times more energy efficient than the SRAM neural cores.

The most recent results for memristor based neuromorphic systems can be seen in [14,15]. Work in [14] did not explain how negative synaptic weights will be implemented in their single memristor per synapse design. Such neurons, capable to represent only positive weights, have very limited capability. No detail programming technique was described. Liu et al. [15] examined memristor based neural networks utilizing four memristors per synapse where the proposed systems utilize two memristors per synapse. They considered deployment of the memristor based system as neural accelerator with a RISC system while proposed systems are standalone embedded processing architectures (which process data directly coming from a 3-D stacked sensor chip).

The rest of the paper is organized as follows: section II describes the overall architecture and the digital neural core design. Section III describes the memristor neural core design and the programming approach. Sections IV and V describe our experimental setup, results respectively. Finally, in section VIII we conclude our work.

## II. MULTICORE ARCHITECTURE

In this study we have examined applications implemented using multi-layer neural networks [16]. Each neuron in such neural network performs two types of operations, (i) a dot product of the inputs $x_1,...,x_n$ and the weights $w_1,...,w_n$, and (ii) the evaluation of an activation function. These operations are shown in Eq. (1) and (2). In a multi-layer feed forward neural network, a nonlinear activation function is desired (e.g. $f(x) = tan^{-1}(x)$).

$$DP_j = \sum_i W_{i,j} x_i \qquad (1)$$
$$y_j = f(DP_j) \qquad (2)$$

Multicore architectures are widely used to exploit task level parallelism. We assumed a multicore neural architecture, for neural network applications, as shown in Fig. 1, with an on-chip routing network to connect the cores. This system processes data directly coming from the sensor chip which is residing on top of the system utilizing 3-D staking technology [17] (see Fig. 2).

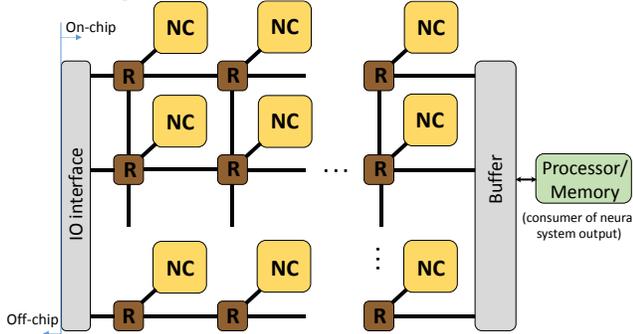

Fig. 1. Proposed multicore system with several neural cores (NC) connected through a 2-D mesh routing network. (R: router).

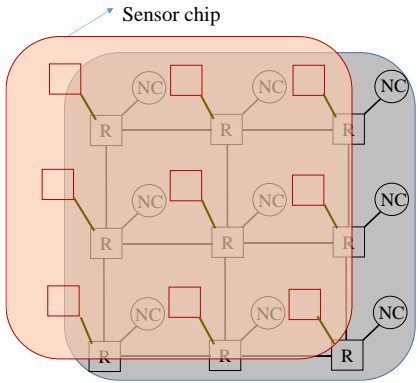

Fig. 2. Sensor chip on top of the neural processing system is shown in this figure. Input data are transferred to the neural chip utilizing through silicon via (TSV).

### A. SRAM Digital Neural Core

Fig. 3 shows a block diagram of the digital neural core. Each core processes a collection of $N$ neurons, with each neuron having up to $M$ input axons. The neuron synaptic weights ($W_{i,j}$) are stored in a memory array. These synaptic values are multiplied with the pre-synaptic input values ($x_i$) and are summed into an accumulator. Once the final neuron outputs are generated, they are sent to other cores for processing, after going through a look-up table implementing an activation function. Input and output buffers store the pre-synaptic inputs and post-synaptic outputs respectively. The memory array shown in Fig. 3 can be developed using several different memory technologies. In this study, we assumed it was implemented using SRAM. In the rest of the paper the system in Fig. 3 will be referred as digital system or SRAM system.

We utilize 8 bits to store a synaptic weight in the SRAM neural core. Every neuron input and output are also 8 bits wide. Inputs belonging to an input pattern (vector) are evaluated one by one. When one input (one component from the input vector) is applied to the core, all the neurons in the core access the corresponding synapses, multiply them with the input, and sum the product in the accumulators simultaneously. In a single SRAM core we have one lookup table for implementing the nonlinear activation function to keep the area and power overhead lower. In this system execution and routing are overlapped. When the core is executing input pattern $n$, it is sending outputs for pattern $n-1$ through the routing network serially (8 bits at a time). As a result, a single lookup table per core is sufficient to evaluate neuron activation function. Section V, subsection B examines determination of near optimum SRAM neural core size.

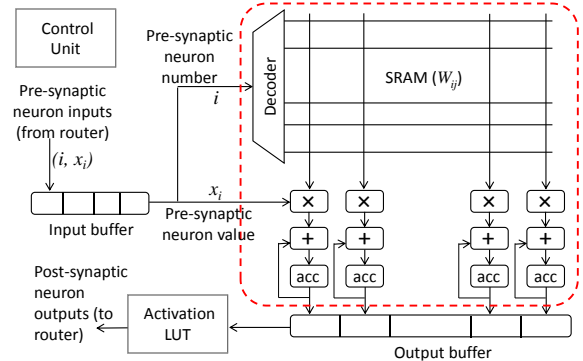

Fig. 3. Proposed digital neural core architecture.

### B. On-chip Routing

An on-chip routing network is needed to transfer neuron outputs among cores in a multicore system. In a feed-forward neural network, the outputs of a neuron layer are sent to the following layer after every iteration (as opposed to a spiking network, where outputs are sent only if a neuron fires). This means that the communication between neurons is deterministic and hence a static routing network can be used for the core-to-core communications. In this study, we assumed a static routing network as this would be lower power consuming than a dynamic routing network. The network is statically time multiplexed between cores for exchanging multiple neuron outputs.

SRAM based static routing is utilized to facilitate re-programmability in the switches [18]. Fig. 4 shows the routing switch design. Note that the switch allows outputs from a core to be routed back into the core to implement recurrent networks or multi-layer networks all within the same core.

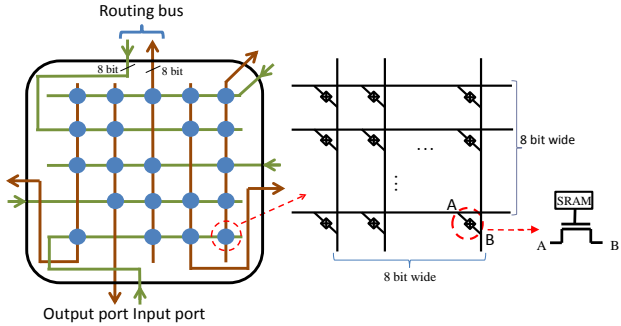

Fig. 4. SRAM based static routing switch. Each blue circle in the left part of the figure represents the 8x8 SRAM based switch shown in the middle (assuming a 8-bit network bus).

## C. IO

The proposed architecture will process data directly coming from sensors in a 3D stack. In the context of embedded systems (e.g., smartphones), the need to go from sensors to memory, and back to the processor, is a major source of energy inefficiency [19]. The neural accelerator has the potential to remove that memory energy bottleneck by introducing sophisticated processing at the level of the sensors, i.e., accelerators between sensors and processors. It can also significantly reduce the amount of data sent to the processor and/or the memory by acting as a preprocessor. For example some applications might require the location of a particular object in a large frame to do additional processing. The neural processing system will identify the target object in the image and will send only the position of the object to the processor for more elaborated processing. The outputs generated by the neural system will be sent to the processor memory for further processing or the processor will read those outputs directly from on-chip buffer adjacent to the neural system (see Fig. 1). In this paper we are particularly focusing on the design and impact of the neural processing system.

## III. MEMRISTOR NEURAL CORE

### A. Memristor Based Neuron Circuit

The schematic in Fig. 5 shows the memristor based neuron circuit used in this study. This example has three inputs, with each input represented by a pair of analog voltages (each one is of opposite polarity of the other). Additionally, two memristors are used to represent a synapse. If the conductance of the memristor connected with input signal is greater than the conductance of the memristor connected with the corresponding inverted signal, then the pair of memristors represent a positive weight (otherwise they represent a negative weight). The output of the inverter pair at the bottom of the circuit represents the neuron output. Power rails of the inverters are taken as $V_{DD}=1$ V and $V_{SS}=-1$ V. In an ideal case, the potential at the first inverter input ($DP_j$) is given by

$$DP_j = \frac{A(\sigma_{A+} - \sigma_{A-}) + \cdots + \beta(\sigma_{\beta+} - \sigma_{\beta-})}{\sigma_{A+} + \sigma_{A-} \cdots + \sigma_{\beta+} + \sigma_{\beta-}} \quad (3)$$

where the conductance of the memristors of Fig. 5 from top to bottom are $\sigma_{A+}$, $\sigma_{A-}$, …, $\sigma_{\beta+}$, $\sigma_{\beta-}$. Eq. (3) indicates that this circuit is essentially carrying out a set of multiply-add operations in parallel in the analog domain. Our experimental evaluations consider memristor crossbar wire resistance through SPICE simulation.

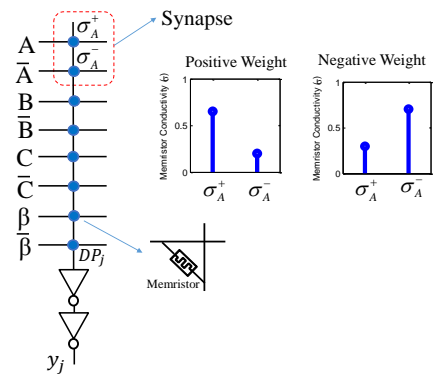

Fig. 5. Circuit diagram for a single memristor-based neuron.

### B. Multi-layer Neural Network

Implementation of a nonlinear separator requires a multi-layer neural network. In a multi-layer feed-forward neural network, as shown in Fig. 6(a), all the neurons in a layer utilize the same set of inputs. Thus the memristor neuron circuit in Fig. 5 can be replicated using memristor crossbars as shown in Fig. 6(b). A layer of memristor based neurons are processed in single cycle in parallel. Each layer of the neurons in Fig 6(a) is replicated using a separate memristor crossbar in Fig 6(b).

The advantage of memristor crossbar based neuron circuit in Fig. 6(b) is that all computations related to a single neural network layer can be evaluated in one step by the crossbar. Additionally computing in the analog domain eliminates the need for multipliers, adders, and accumulators. This leads to area and power efficiency along with high computation throughput. The non-volatile nature of memristors allows the circuit to be turned off when not in use, thus reducing the static power consumption.

A nonlinear activation function is typically used to generate a neuron's output (given by Eq 2). To enable parallel operations in the circuit in Fig. 6(b), the activation function circuit has to be reproduced for each neuron. To keep the area of the neuron circuit low, a threshold activation function is used in this study, consisting of a pair of inverters, as shown in Fig.

5. More complex activation functions (such as sigmoid) would require a costly analog-to-digital converter to be placed at each neuron output, adding significant area overhead (about 3000 transistors per neuron for 8 bits neuron output) and power overheads.

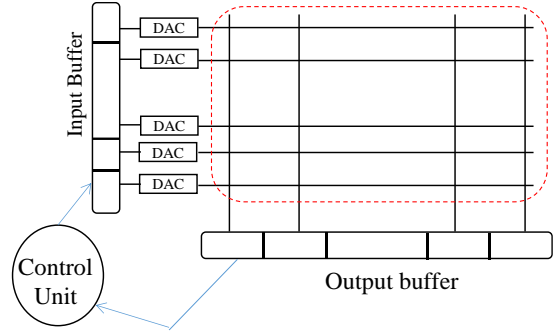

Fig. 8. Neural core having DACs for processing the first neuron layer of a network.

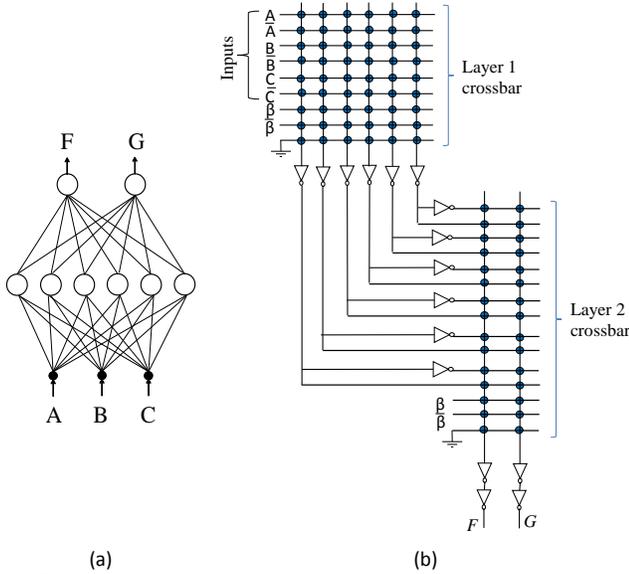

(a)             (b)

Fig. 6. (a) Two layer network and (b) memristor crossbar based implementation of the neural network.

### C. Memristor Neural Core Design

Fig. 7 shows the memristor based single neural core architecture. It is consisting of a memristor crossbar of certain capacity, input and output buffers, and a control unit. The control unit will manage the input and output buffers and will interact with the corresponding routing switch associated with the core. The control unit will be implemented as a finite state machine and thus will be of low overhead. This is very similar to the control unit in the digital SRAM core. Near optimum core size is examined in section V, subsection B.

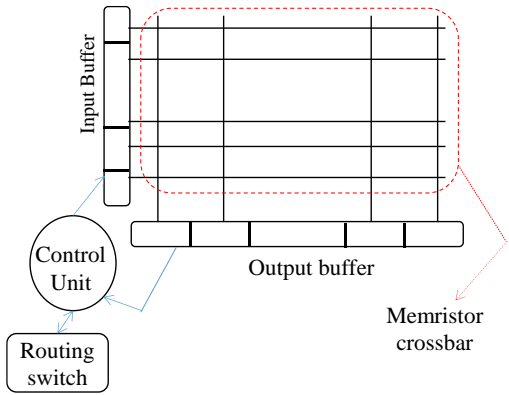

Fig. 7. Memristor crossbar based neural core architecture.

As the inputs are coming from a different sensor chip to the neural processing chip, inputs should be in digital form for ease of transmission. In this paper we are assuming each input is represented by 8 bits. Before applying inputs to the memristor neuron circuit, these need to be converted to analog form. Fig. 8 shows the neural core implementing first hidden layer neurons of the neural network which utilizes digital to analog converters (DAC). Neural cores of both types (having and not having DAC) are distributed uniformly over the chip.

### D. Programming of the Memristor Cores

To implement a desired functionality, a memristor crossbar neural network needs to be programmed. We utilized ex-situ training for memristor based neural networks. The objective of the off-chip (ex-situ) training process is to set each memristor in the crossbar to a specific resistance within the programmable range. Programming memristor crossbars is challenging because they have a significant amount of variation present between devices. This means that identical voltage pulses may not induce identical amounts of resistance change in different memristors within a crossbar. Multiple pulses may be required to set the memristors to a target state while reading the device state between these write pulses. This is essentially a feedback write process which requires being able to read the resistance of each individual memristor.

In the off-chip training process, reading each memristor resistance level is challenging due to sneak-paths in the crossbar. Placing a transistor at each cross-point of a crossbar will ensure that only the resistance of the target memristor is impacting the column voltage during the programming process. Such crossbar is known as 1T1M crossbar. Fig. 9 shows that when reading the 1T1M system enabling $S_{R1}$, the only current path available in the first column is through $M_{11}$ (memristor at the first row and first column) and the known resistance $R_{S1}$.

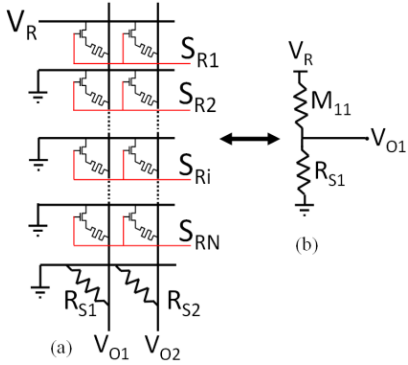

Fig. 9. Reading a memristor from the 1T1M crossbar.

For the memristor neural cores we assumed that the off-chip programming process will be coordinated by an off-chip system and so added only the necessary hardware needed to allow the external system to program the crossbar. These include a single A-to-D converter, a set of transistors, a set of buffers and control circuits per core. The use of a single A-to-D converter per core will serialize the programming process for each core. We assume, this is not a problem as this system will be programmed once and then deployed for use. An earlier study shows that 7 bits of precision is achievable from a single memristor [20]. Memristor based neuron circuits are utilizing two memristors per synapse which will provide combined synaptic weight precision of about 8 bits.

## IV. EVALUATION OF PROPOSED ARCHITECTURES

### A. Memristor Model

Simulation of the memristor device used an accurate model of the device published in [21]. The memristor device simulated in this paper was published in [22] and the switching characteristics for the model are displayed in Fig. 10. This device was chosen for its high minimum resistance value and large resistance ratio. According to the data presented in [22] this device has a minimum resistance of 125 kΩ, a resistance ratio of 1000, and the full resistance range of the device can be switched in 80 ns applying 4.25 V across the device.

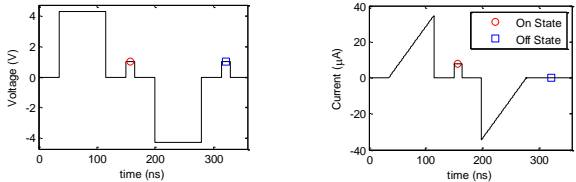

Fig. 10. Simulation results displaying the input voltage and current waveforms for the memristor model [21] that was based on the device in [22]. The following parameter values were used in the model to obtain this result: $V_p$=4V, $V_n$=4V, $A_p$=816000, $A_n$=816000, $x_p$=0.9897, $x_n$=0.9897, $\alpha_p$=0.2, $\alpha_n$=0.2, $a_1$=1.6×10$^{-4}$, $a_2$=1.6×10$^{-4}$, $b$=0.05, $x_0$=0.001.

### B. Application Description

We have selected the following five applications for our system level evaluations: edge detection, motion estimation, deep networks, object recognition, and optical character recognition (OCR). These are described in detail below.

**Edge** (edge detection): We implemented Sobel edge detection algorithm that takes 3x3 pixels as input to generate one output pixel. The application was implemented using convolution operation (not using neural network) on the RISC processing cores. This makes sure that the best algorithm was used for the RISC system. In the SRAM neural cores, the algorithm was approximated using a neural network with configuration 9→20→1 (9 inputs, 20 neurons in hidden layer and 1 output neuron). For the memristor systems we utilized four neural networks of configurations 9→20→15, 24→20→15, 15→10→4, and 15→10→4. These extra networks were needed to generate the multi-bit outputs for the application.

**Deep:** Deep neural networks have become popular for face, object and pattern recognition tasks. We developed a small scale deep network to process images from the MNIST dataset [23]. This dataset contains 60,000 images of handwritten digits, with each image consisting of 28×28 grayscale pixels. We utilized a neural network with configuration 784 →200→100→10. The network was trained with 50000 images from the dataset. The same network was used in both the RISC processor and the neural processors.

**Motion:** Two images are compared to determine the degree of motion within the images. The algorithm estimates the degree of motion in increments of 5% from 0% to 50%. For an $m \times n$ frame size, to detect motion we determined pixel deviations in the 8×8 grids and accumulated those deviations. For the memristor system, we utilized neural networks of configuration 64(2→1), 64→10, and 20→10. For the SRAM system, utilized neural network configurations are 64(2→1), 64→1, and 2→1. The SRAM system network is different as it has multi-bit outputs. The application was implemented for the RISC system on the basis of calculation of pixel deviation (not in neural network form).

**Object Recognition:** We have examined an object recognition task on the CIFAR-10 dataset [24]. This dataset consists of 60,000 color images of size 32×32 belonging to 10 classes, including airplanes, automobiles, birds, cats, deer, dogs, frogs, horses, ships, and trucks. For the desired classification we utilized a two layer neural network with configuration 3072→100→10.

**OCR**: The Optical Character Recognition application deals with the recognition and classification of printed characters. The neural network was trained using the Chars74K dataset [25], consisting of 128×128 pixel images. We subsampled the character images and used 50×50 images in our experiment. We utilized a network of configuration 2500→60→26 on all the systems.

### C. Mapping Neural Networks to Cores

The neural hardware are not able to time multiplex neurons as their synaptic weights are stored directly within the neural circuits. Hence a neural network's structure may need to be modified to fit into the neural cores. In cases where the networks are significantly smaller than the neural core memory, multiple neural layers are mapped to a core. In this case, the layers execute in a pipelined manner, where the outputs of layer 1 are fed back into layer 2 on the same core through the adjacent routing switch.

When a network layer is too large to fit into a core (either because it needs too many inputs or it has too many neurons), the network layer is split amongst multiple cores. Splitting a layer across multiple cores due to large number of output neurons is trivial. When there are too many inputs per neuron for a core, each neuron is split into multiple smaller neurons and then combined through a higher level neuron as shown in Fig. 11. When splitting a neuron, the network needs to be trained to account for the new network topology. As the network topology is determined prior to training (based on the neural hardware architecture), the split neuron weights are trained correctly.

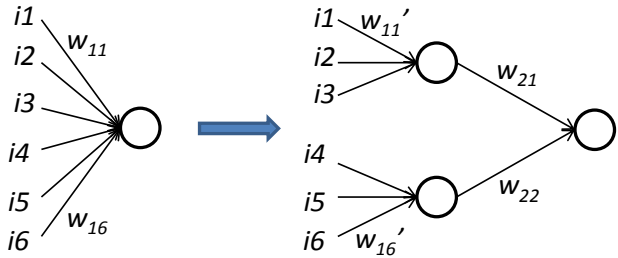

Fig. 11. Splitting a neuron into multiple smaller neurons.

### D. Area Power Calculations

**RISC core:** We have compared power, performance of memristor based systems with a traditional RISC system. The examined ARM processor is a single issue, inorder 6 stage pipelined system and operates at 1 GHz clock. L1 instruction cache size is 16 kB and L1 data cache is 16 kB. We assumed main memory access latency 1 cycle to mimic prefetching from 3D stacked main memory. The reason behind choosing the single issue inorder processor over a superscalar out-of-order process is to achieve overall power efficiency for the real time applications. A complex out-of-order processor provides more throughput over a simple single issue inorder processor, but consumes significantly more power. We assumed all systems used a 45nm process for this study. We have obtained area and power of the RISC core utilizing McPat [26]. Area of the core is 0.524 mm$^2$ and it consumes 0.087 W power. Performance of the RISC core was evaluated utilizing SimpleScalar [27] simulator.

**Specialized cores:** The area, power, and timing of the SRAM array were calculated using CACTI [28] with the low operating power transistor option utilized. Components of a typical cache that would not be needed in the neural core (such as the tag array and tag comparator) were not included in the calculations. The power of the basic components (such as multipliers, adders, and registers) were determined through SPICE simulations.

The routing link power was calculated using Orion [29] (assuming 8 bits per link). A frequency of 200 MHz was assumed for the digital system to keep power consumption low. Off-chip IO energy was also considered as described in Section II.

Data transfer energy via TSV was assumed to be 0.05 pJ/bit [30].

For the memristor cores, detailed SPICE simulations were used for power and timing calculations of the analog circuits (drivers, crossbar, and activation function circuits). These simulations considered the wire resistance and capacitance within the crossbar as well. The results show that the crossbar required 10 ns for processing. As the memristor crossbars evaluate all neurons in one step, the majority of time in these systems is spent in transferring neuron outputs between cores through the routing network. We assumed that routing would run at 200 MHz clock resulting in two cycles needed for crossbar processing.

## V. RESULTS

### A. Bit Width and Activation Function

The memristor based system implements threshold activation function using a pair of inverters. In the SRAM architecture, the activation function is implemented using a lookup table. Since the neuron outputs are transmitted over the on-chip router serially in the SRAM architecture, only one lookup table is needed per core. The area and power overhead of using the lookup table was 1% and 0.3% respectively for a 256×128 (inputs×outputs) digital core.

The bit width of the synaptic weights and the neuron activation function, $f$ in Eq. (2), will impact the output accuracy of the computations. As shown in Fig. 12, a synaptic bit width of 8 bits results in an average loss in accuracy of less than 1% and 3% for the sigmoid and threshold activation functions respectively. Therefore both the memristor based and the SRAM architectures utilized the 8 bit precision for weights.

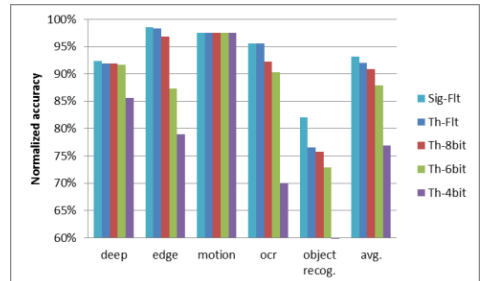

Fig 12. Error for different precisions (with the same number of neurons). Here Sig denotes for sigmoid, Flt for floating point, and Th for threshold.

### B. Design Space Exploration of Neural Cores

For each of the three neural core types, we varied the memory/crossbar array size to examine the impact on area and power as shown in Figs 13, and 14. Each of the applications was mapped to the multicore system based on the method in Section IV, subsection C. Image sizes of 2500×2500 were used in this design space exploration.

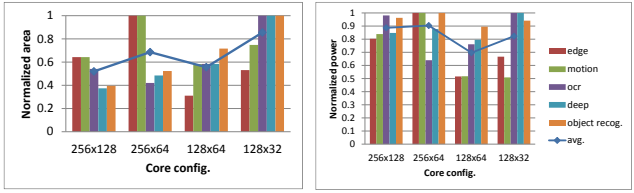
Fig. 13. 1T1M system area and power.

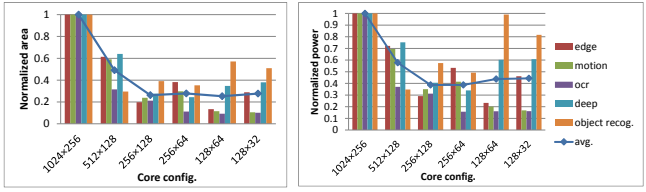
Fig. 14. Digital system area and power.

For the memristor based systems, we picked the 128×64 (inputs × outputs) memristor crossbar core configuration as this had the lowest average normalized area and power for the different applications. For the digital systems, the optimum size was 256×128 synapses in the on-core memory array; this corresponds to a memory size of 256×128 bytes (8 bits per synapse). Due to the 8-bit outputs, the lookup table needed 256 bytes of memory to implement the activation function. Table I show area and power of different cores.

Table I: Area and power of different cores.

|  | Area (mm$^2$) | Total power (mW) | Leakage power (mW) | Core processing time (sec.) |
| --- | --- | --- | --- | --- |
| RISC | 0.524 | 87 | 54 | 3.97×10$^{-5}$ (1 neuron, 784 synapse) |
| Digital | 0.208 | 24.2 | 6.94 | 1.28×10$^{-6}$ (128 neuron, 256 synapse/neuron) |
| 1T1M | 0.0082 | 0.0888 | 0.0118 | 9×10$^{-8}$ (64 neuron, 128 synapse/neuron) |

## C. Results for Real Time Applications

To compare the two neural architectures against RISC processing cores, we examined the processing of real time application loads:
- Deep network/character recognition: process 100,000 characters per second. For the deep network, inputs are 28×28 pixel handwritten digits, while for character recognition, these are 50×50 printed characters.
- Edge detection and motion estimation: process an 1280×1080 image stream at 60 frames per second.

Tables II to VI show the number of cores, area, and power for the different architectures to process these applications with the specified real time processing requirements. The results show that the digital neural processor is about 14 to 952 times more efficient than the RISC cores, while the memristor architectures are about 5,641 to 187,064 times more efficient. We assumed that during the idle time, the memristor neural cores would not consume significant static power.

Table II: Deep Network

|  | Number of cores | Area (mm$^2$) | Power (mW) | Power efficiency over RISC |
| --- | --- | --- | --- | --- |
| RISC | 902 | 472.65 | 78474.00 | 1 |
| Digital | 9 | 1.88 | 82.40 | 952 |
| 1T1M | 31 | 0.25 | 0.42 | 187064 |

Table III: Edge Detection

|  | Number of cores | Area (mm$^2$) | Power (mW) | Power efficiency over RISC |
| --- | --- | --- | --- | --- |
| RISC | 240 | 125.76 | 20880.00 | 1 |
| Digital | 18 | 3.75 | 433.16 | 48 |
| 1T1M | 16 | 0.13 | 1.41 | 14813 |

Table IV: Motion Estimation

|  | Number of cores | Area (mm$^2$) | Power (mW) | Power efficiency over RISC |
| --- | --- | --- | --- | --- |
| RISC | 7 | 3.67 | 609.00 | 1 |
| Digital | 2 | 0.42 | 42.57 | 14 |
| 1T1M | 2 | 0.02 | 0.11 | 5641 |

Table V: Object Recognition

|  | Number of cores | Area (mm$^2$) | Power (mW) | Power efficiency over RISC |
| --- | --- | --- | --- | --- |
| RISC | 1358 | 711.59 | 118146.00 | 1 |
| Digital | 17 | 3.54 | 148.55 | 795 |
| 1T1M | 68 | 0.56 | 0.94 | 125430 |

Table VI: Optical Character Recognition

|  | Number of cores | Area (mm$^2$) | Power (mW) | Power efficiency over RISC |
| --- | --- | --- | --- | --- |
| RISC | 825 | 432.30 | 71775.00 | 1 |
| Digital | 13 | 2.71 | 119.08 | 603 |
| 1T1M | 31 | 0.25 | 0.49 | 147012 |

In the RISC system, edge detection and motion detection applications are executed in traditional algorithmic procedure (not in neural network form) to make sure that the best algorithms are executed in this system. In the specialized architectures, these two applications are executed in neural network form. Neural network representations of these two applications increase the number of operations to be performed compared to the original RISC operations. As a result power efficiencies for these two applications in the specialized architectures are not as high as the actual neural network applications (deep, object recog., OCR).

## VI. CONCLUSION

In this paper we have performed full system evaluation of the multicore systems based on memristor neural cores. On-chip routing requirements, IO interface were examined for these systems. We have performed design space exploration of specialized neural cores and determined optimum neural core configurations. Synaptic memory accesses in the digital system consume significant portion of the overall system powers. In the memristor based systems data processing takes place at the physical location of the data. Parallel analog operation of the memristor crossbar does not require adders and multipliers to perform the neuron operations. Specialized architectures for neural networks provide higher throughput over RISC architecture. Furthermore, parallel analog operation of the memristor based systems provide dramatic throughput and power efficiencies over digital systems.